\documentclass[aps,prl,superscriptaddress,twocolumn]{revtex4}
\usepackage{graphicx}

\begin{document}

\title{A combinatorial approach to metamaterials discovery}

\author{E. Plum}
\affiliation{Optoelectronics Research Centre and Centre for Photonic
Metamaterials, University of Southampton, Southampton, SO17 1BJ, UK}
\email{erp@orc.soton.ac.uk}

\author{K. Tanaka}
\affiliation{Optoelectronics Research Centre and Centre for Photonic
Metamaterials, University of Southampton, Southampton, SO17 1BJ, UK}
\affiliation{Sony Corporation, Shinagawa-ku, Tokyo, 141-0001, Japan}

\author{W. T. Chen}
\affiliation{Department of Physics, National Taiwan University,
Taipei, 10617, Taiwan}

\author{V. A. Fedotov}
\affiliation{Optoelectronics Research Centre and Centre for Photonic
Metamaterials, University of Southampton, Southampton, SO17 1BJ, UK}

\author{D. P. Tsai}
\affiliation{Department of Physics, National Taiwan University,
Taipei, 10617, Taiwan}

\author{N. I. Zheludev}
\affiliation{Optoelectronics Research Centre and Centre for Photonic
Metamaterials, University of Southampton, Southampton, SO17 1BJ, UK}

\begin{abstract}
We report a high through-put combinatorial approach to photonic
metamaterial optimization. The new approach is based on parallel
synthesis and consecutive optical characterization of large numbers
of spatially addressable nano-fabricated metamaterial samples
(libraries) with quasi-continuous variation of design parameters
under real manufacturing conditions. We illustrate this method for
Fano-resonance plasmonic nanostructures arriving at explicit recipes
for high quality factors needed for switching and sensing
applications.
\end{abstract}


\maketitle

Some fifteen years ago a paper reporting a combinatorial approach to
materials discoveries revolutionized materials research and other
disciplines such as chemistry and pharmacology by demonstrating a
method for high-throughput parallel synthesis and analysis of novel
artificial chemical compounds \cite{Xiang_CombinatorialMaterials,
NatMat_2004_Combinatorial}. Here we report on the first application
of the combinatorial approach to discovery of electromagnetic
metamaterials and their optimization. Electromagnetic metamaterials
are manmade media with all sorts of unusual functionalities that can
be achieved by artificial structuring on a smaller length scale than
that of the external stimulus. We apply the combinatorial approach
to the optimization of an important class of plasmonic metamaterials
supporting Fano resonances \cite{PRL_Fedotov_2007_TrappedModes} that
recently became a prime platform for new switching, gain and sensing
applications, slow light and polarization control devices (for a
review see \cite{NatMat_2010_FanoResonance}). These metamaterials
are arrays of asymmetric split ring metallic wire resonators or
complementary arrays of ring slits in a metal film. Their
application potential depends on the presence and characteristics of
reflection and transmission peaks associated with a particular mode
of electromagnetic excitation that is weakly coupled to free-space
(closed or trapped mode). The characteristics of this resonance are
very sensitive to the environment and are responsive to its changes
which makes them a prime choice for active metamaterial
applications. These resonances can be tailored by design: their
spectral position, width and depth depend on many parameters, most
notably on the type of metal used for their fabrication, size of the
ring and characteristics of the split
\cite{APB_Rockstuhl_2006_SRR_resonances,
OE_2010_Atwater_SymBreaking, NatMat_Liu_2009_EIT,
JOpt_Khardikov_2010_LightTrapping}. Optimization of resonance
characteristics of these metamaterial structures is relatively
straightforward in the microwave and terahertz parts of the spectrum
using numerical electromagnetic modeling tools. For the optical part
of the spectrum when the meta-molecules should be structured on the
nanoscale numerical modeling becomes considerably less reliable for
two reasons. First, accurate reproduction of the idealized design
parameters is not possible due to limitations of the nanofabrication
technologies, in particular sharp edges and ideal vertical cuts are
not possible to achieve while the surface of metals has roughness on
a scale comparable with the smallest features of the design. Second,
electromagnetic material properties used in calculations, in
particular those of nanostructured metals are not known to the
necessary accuracy and are normally replaced by those of bulk media.
However, they can be drastically different from those of bulk metals
due to the higher role of surface electron and defect scattering in
granulated and nanostructured metals. The unsatisfactory character
of this substitute is well recognized by the metamaterials research
community where in some calculations arbitrarily corrected values of
bulk dielectric parameters are used for nanostructured metals to
achieve a better fit to experimental data.

\begin{figure*}[tb!]
\includegraphics[width=135mm]{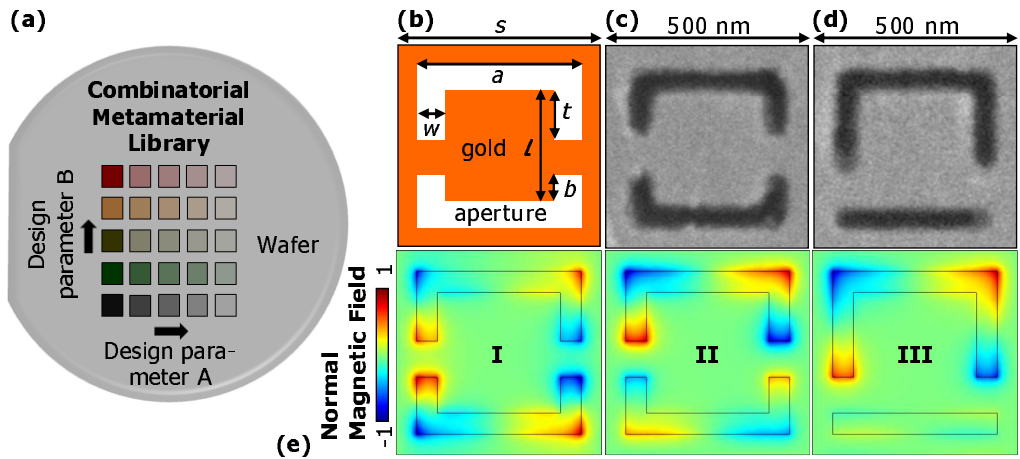}
\caption{\label{fig1}(Color online) \textbf{Combinatorial library,
meta-molecules and resonant modes.} (a) Schematic of a combinatorial
metamaterial library. (b) Schematic of the metamaterial unit cell.
(c) and (d) show scanning electron micrographs of metamaterial unit
cells of size $s=500$~nm with small and large gap position
asymmetries, $\beta=1/7$ and 1. (e) Characteristic \textbf{I}
symmetric and \textbf{II}-\textbf{III} antisymmetric modes of
excitation associated with the absorption resonances for asymmetries
$\beta=1/7$ and 1. The plotted field distributions correspond to
excitation wavelengths of 820, 1025 and 1200~nm, respectively.}
\end{figure*}

The combinatorial approached presented below aims to address the
difficulty of optimizing photonic metamaterial designs through
computer modeling by resorting to a parallel synthesis and
consecutive optical characterization of a large number of spatially
addressable nanofabricated metamaterial samples [libraries, see Fig.
\ref{fig1}(a)] with quasi-continuous variation of design parameters.
We also compare the results of combinatorial optimization with those
of numerical modeling.

We studied metamaterials consisting of arrays of square split-ring
aperture meta-molecules in the field of parameters defined by the
overall size $s$ of the metamaterial unit cell and the asymmetry of
the split [see Fig. \ref{fig1}(b)]. Here $s\times s$ is the overall
size of the square unit cell while in all cases the split ring
aperture's outer and inner sides $a$ and $l$ are 100~nm and 200~nm
smaller than the unit cell, respectively. The design slit width $w$
was 50~nm and parameters $t$ and $b$ define the asymmetry of the
split and its size. Two libraries of regular $30\times 30
\mu\text{m}^2$ arrays of split ring apertures in a 30~nm thick layer
of gold were fabricated by electron beam lithography on a glass
wafer. In both libraries the unit cell size $s$ was varied from
400~nm to 500~nm in steps of 25~nm. In library A parameter $b=0$ was
fixed and a gap size asymmetry $\alpha=t/l$ ranging from 0.1 to 0.7
in steps of 0.1 was introduced. In library B the gap size was fixed
at $0.3l$ and a gap position asymmetry $\beta=(t-b)/(t+b)$ ranging
from 0 to 1 in steps of 1/7 was introduced.

The transmission and reflection characteristics of these photonic
metamaterials were measured in the spectral range from 800 to
2000~nm for waves polarized parallel to $t$ using a
microspectrophotometer. These properties were also simulated with a
full three-dimensional Maxwell finite element method solver in the
frequency domain using the Drude model for the dielectric constant
of gold \cite{PRB_Johnson_1972_MetalParameters}.

\begin{figure*}[tb!]
\includegraphics[width=170mm]{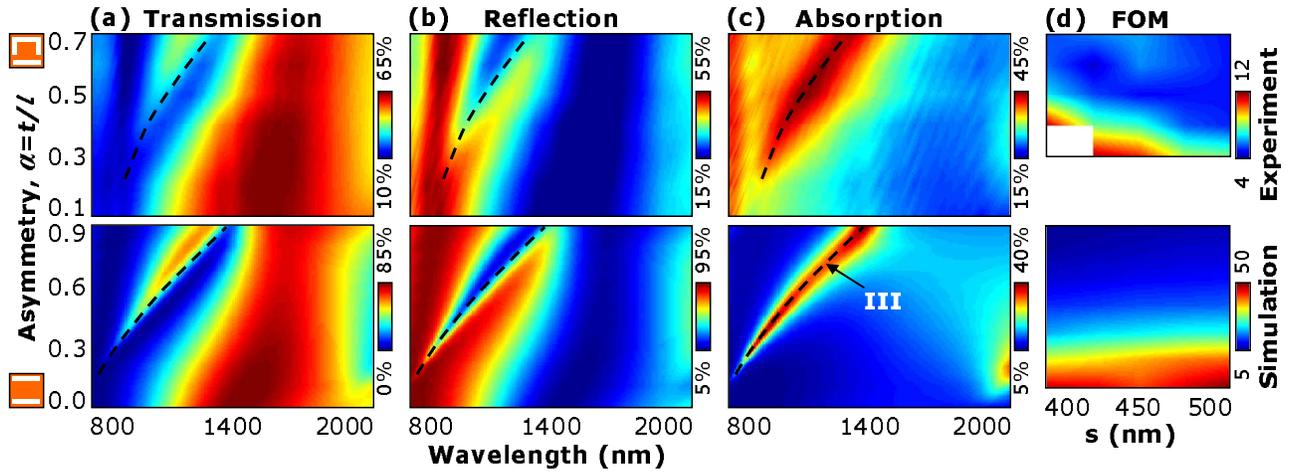}
\caption{\label{fig_gapsize}(Color online) \textbf{Library A - gap
size asymmetry} $\alpha=t/l$, where parameters $l=s-200$~nm and
$b=0$ remain fixed. (a) Transmission, (b) reflection and (c)
absorption spectra as a function of symmetry breaking $\alpha$ for
unit cell size $s=500$~nm. The dashed line corresponds to the
absorption maximum. (d) Combinatorial maps of resonance sharpness as
a function of unit cell size $s$ versus gap size asymmetry $\alpha$.
Here the resonance width is measured as a figure of merit,
$FOM=\lambda_0/\Delta\lambda$, defined as the resonance wavelength
$\lambda_0$ of the absorption peak divided by its full width half
maximum $\Delta\lambda$. Note that the experiments (top) cover a
slightly smaller range of asymmetries than the simulations
(bottom).}
\end{figure*}

Figure \ref{fig_gapsize}(a)-(c) illustrates for a fixed
meta-molecule size of $s=500$~nm how the metamaterial's
transmission, reflection and absorption properties depend on the gap
size asymmetry $\alpha=t/l$ when $b=0$. In the symmetric case the
metamaterial has broad reflection and transmission maxima around
800~nm and 1400~nm respectively, while the absorption spectrum is
featureless. Symmetry breaking, $\alpha>0$, leads to the appearance
of a Fano resonance near 900~nm which is associated with a narrow
peak in absorption. Furthermore, at slightly shorter wavelengths, a
window of transparency opens up, which may be discussed in terms of
electromagnetically induced transparency
\cite{Papasimakis_EIT_2009}. Larger symmetry breaking leads to
broadening of the absorption resonance, which red-shifts due to the
increasing size of the top aperture.

\begin{figure*}[tb!]
\includegraphics[width=170mm]{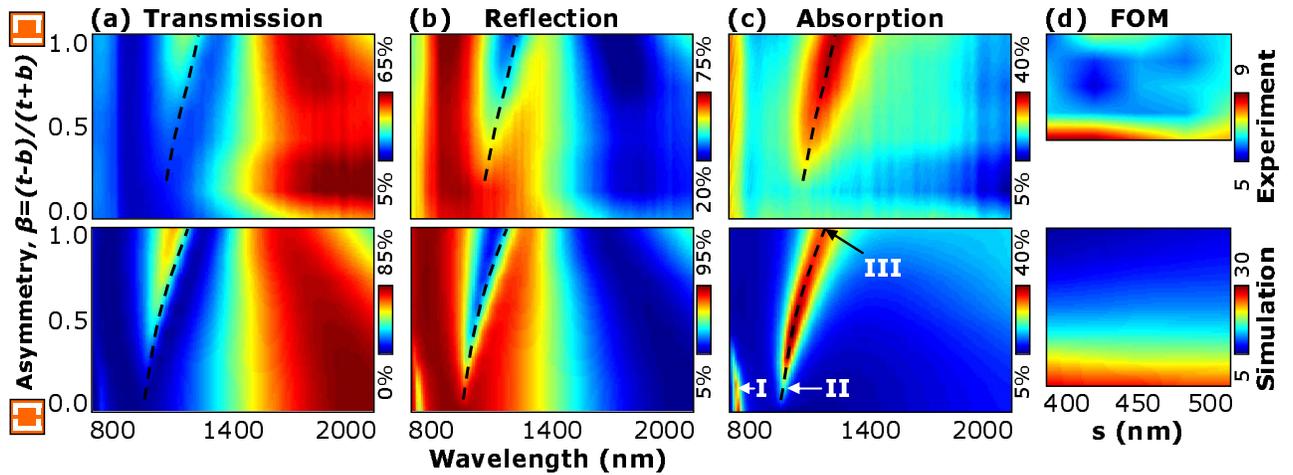}
\caption{\label{fig_gapposition}(Color online) \textbf{Library B -
gap position asymmetry} $\beta=(t-b)/(t+b)$, where parameters
$l=s-200$~nm and $t+b=0.7l$ remain fixed. (a) Transmission, (b)
reflection and (c) absorption spectra as a function of symmetry
breaking $\beta$ for unit cell size $s=500$~nm. The dashed line
corresponds to the absorption maximum. (d) Combinatorial maps of
resonance sharpness $FOM$ as a function of unit cell size $s$ versus
gap position asymmetry $\beta$. The resonant modes of excitation
corresponding to \textbf{I}-\textbf{III} are shown in
Fig.~\ref{fig1}(e).}
\end{figure*}

Figure \ref{fig_gapposition}(a)-(c) shows how the metamaterial
transmission, reflection and absorption depend on the gap position
asymmetry $\beta=(t-b)/(t+b)$ for a fixed 90~nm gap and a unit cell
size of $s=500$~nm. Similarly to the gap size asymmetry discussed
above, a gap position asymmetry $\beta>0$ leads to the appearance of
a Fano resonance at 1000~nm which broadens and red-shifts with
increasing asymmetry. Intriguingly, simulations reveal an additional
narrow absorption resonance at 830~nm, which should be supported by
the symmetric structure and vanishes with increasing asymmetry
$\beta$. The fundamentally different nature of these resonances
becomes clear when comparing the associated modes of excitation
shown in Fig. \ref{fig1}(e). The symmetric structure supports a
symmetric mode \textbf{I}, where top and bottom apertures interact
through in-phase magnetic fields. Symmetry breaking allows the
excitation of an anti-symmetric mode \textbf{II}, where the
apertures of each split ring interact through anti-phase magnetic
fields. These modes can be understood as the hybrid modes of two
coupled resonators (apertures): \textbf{I} being the higher energy
anti-bonding and \textbf{II} being the lower energy bonding mode
\cite{FrontPhysChina_Liu_2010_Hybridization}. With increasing
symmetry breaking the bonding mode \textbf{II} becomes dominated by
the longer aperture, see \textbf{III}.

As discussed in the introduction, narrow Fano resonances have a wide
range of applications. To quantify the width of the bonding
resonance we introduce a figure of merit defined as
$FOM=\lambda_0/\Delta\lambda$, where $\lambda_0$  is the resonant
wavelength and $\Delta\lambda$  is the full width half maximum of
the absorption peak. As illustrated by Figs. \ref{fig_gapsize}(d)
and \ref{fig_gapposition}(d), for both types of symmetry breaking
smaller asymmetries lead to narrower Fano resonances, while the
resonance width does not strongly depend on the meta-molecule size,
which controls the spectral position of the resonance.

As illustrated by Figs. \ref{fig_gapsize} and \ref{fig_gapposition},
numerical results and experiments show similar qualitative behavior,
indicating that simulations do provide useful guidance for the
design and understanding of photonic metamaterials. However,
quantitative comparison shows substantial differences between
experimental and numerical results.

In particular, the spectral position of the measured absorption
resonances is red-shifted by about 80~nm relative to the simulation
results. This shift may be explained by uncertainty regarding the
real part of the dielectric constant of nanoscale volumes of gold at
optical frequencies. Also systematic manufacturing inaccuracies like
the inability to fabricate sharp corners at the nanoscale contribute
to a spectral shift of resonances, see Fig. \ref{fig1}(c)-(d). In
general spectral positions of Fano-type resonances in plasmonic
metamaterials are also extremely sensitive to thin cover layers, or
layers of oxides or contamination, as has recently been demonstrated
by a 3-fold increase in metamaterial transmission resulting from
application of a single layer of graphene
\cite{OE_Papasimakis_2010_Graphene}.

While simulations predict remarkably narrow resonances of up to
$FOM=50$ for small asymmetries around 0.1, the experimentally
observed resonances are broadened $FOM\leq12$ and weakened so much,
that they cannot be clearly identified for asymmetries
$\alpha\leq0.3$ and $\beta\leq3/7$. Small size variations and
manufacturing tolerances in reproducing a perfect square array of
meta-molecules \cite{PRB_Papasimakis_2009_Coherent}, rough gold
surfaces as well as finite array sizes
\cite{PRL_2010_Fedotov_ArraySizes} were shown to lead to resonance
broadening and weakening in real structures, while these factors are
not accounted for in simulations of infinite arrays of identical,
ideally smooth meta-molecules. Additionally, the width of resonances
strongly depends on the imaginary part of the dielectric constant of
gold, which may be larger than the commonly used literature values
\cite{PRB_Johnson_1972_MetalParameters} for nanoscale volumes of
gold in the optical part of the spectrum due to increased rates of
electron scattering. These factors can also explain why the
remarkably narrow anti-bonding resonance associated with the
symmetric mode of excitation has not been observed in real
structures. Furthermore, the above factors can also cause the high
background of absorption observed in our real metamaterial samples.

For practical applications a trade-off between narrow Fano
resonances (small asymmetry) and strong Fano resonances (large
asymmetry) must be considered. The optimal design parameters
(i.e.~asymmetry) will always depend on the accuracy of the available
nanostructuring technology and can be found through the high
through-put screening method demonstrated here. In our case
asymmetries around $\alpha, \beta\simeq0.4-0.5$ are the best choice
for most sensor and switching applications, which is in stark
contrast to simulations suggesting optimal asymmetries of around
0.1.

In summary we report a high through-put combinatorial approach to
the design and optimization of photonic metamaterials. We show for
Fano resonances in split ring aperture arrays, that the optimal
metamaterial parameters can be found through parallel synthesis and
characterization of material libraries with quasi-continuous
variation of design parameters under real manufacturing conditions.
Being unable to take the inevitable shortcomings of nanoscale
manufacturing into account, simulations can only provide qualitative
insights into metamaterial properties and underlying mechanisms.
Here we have identified in detail how the asymmetry of split ring
apertures controls Fano resonances in photonic metamaterials, which
are relevant to various applications ranging from sensing, switches
and slow-light devices to the ``lasing spaser"
\cite{Zheludev_spaser_2008}.

\begin{acknowledgments}
The authors are grateful to Jun Yu Ou for his assistance in imaging
of metamaterial samples. Financial support of the Engineering and
Physical Sciences Research Council, UK is acknowledged.
\end{acknowledgments}

\end{document}